\documentclass{Rinton-P9x6}

\begin{document}

\title{Phenomenology of Anti-de-Sitter Conformal Field Theory Duality}

\author{Paul H. Frampton}

\address{Institute of Field Physics, Department of Physics and Astronomy,
University of North Carolina, Chapel Hill, NC 27599-3255}


\maketitle

\abstracts{
By use of the AdS/CFT correspondence on orbifolds, models
are derived which can contain the standard model
of particle phenomenology. It will be assumed that the theory becomes
conformally invariant at a renormalization-group
fixed-point in the TeV region. }

\section{Introduction}

It is a privilege to speak at the first High-Energy Physics conference
to be held in Cairo, hopefully the first of many. In this talk,
only an outline of the conformality idea can be described; more detail is
to be found in the references.

\bigskip

In particle phenomenology, the impressive success of the standard theory
based on $SU(3) \times SU(2) \times U(1)$ has naturally led to the question
of how to extend the theory to higher energies? One is necessarily led by
weaknesses and incompleteness in the standard theory. If one extrapolates
the standard theory as it stands one finds (approximate) unification of the
gauge couplings at $\sim 10^{16}$ GeV. But then there is the {\it hierarchy}
problem of how to explain the occurrence of the tiny dimensionless ratio $%
\sim 10^{-14}$ of the weak scale to the unification scale. Inclusion of
gravity leads to a {\it super-hierarchy} problem of the ratio of the weak
scale to the Planck scale, $\sim 10^{18}$ GeV, an even tinier $\sim 10^{-16}$.
Although this is obviously a very important problem about which
conformality by itself is not informative, we shall discuss first the
hierarchy rather than the super-hierarchy.

\bigskip

There are four well-defined approaches to the hierarchy problem:

\begin{itemize}
\item  1. Supersymmetry

\item  2. Technicolor.

\item  3. Extra dimensions.

\item  4. Conformality.
\end{itemize}

\noindent {\it Supersymmetry} has the advantage of rendering the hierarchy
technically natural, that once the hierarchy is put in to the lagrangian it
need not be retuned in perturbation theory. Supersymmetry predicts
superpartners of all the known particles and these are predicted to be at or
below a TeV scale if supersymmetry is related to the electroweak breaking.
Inclusion of such hypothetical states improves the gauge coupling
unification. On the negative side, supersymmetry does not explain the origin
of the hierarchy.

\bigskip

\noindent {\it Technicolor} postulates that the Higgs boson is a composite
of fermion-antifermion bound by a new (technicolor) strong dynamics at or
below the TeV scale. This obviates the hierarchy problem. On the minus side,
no convincing simple model of technicolor has been found.

\bigskip

\noindent {\it Extra dimensions} can have a range as large as $1 ({\rm TeV}%
)^{-1}$ and the gauge coupling unification can happen quite differently than
in only four spacetime dimensions. This replaces the hierarchy problem with
a different fine-tuning question of why the extra dimension is restricted to
a distance corresponding to the weak interaction scale.

\bigskip

\noindent {\it Conformality} is inspired by superstring duality and assumes
that the particle spectrum of the standard model is enriched such that there
is a conformal fixed point of the renormalization group at the TeV scale.
Above this scale the coupling do not run so the hierarchy is nullified.

\bigskip

Conformality is the approach followed in this paper. We shall systematicaly
analyse the compactification of the IIB superstring on $AdS_5 \times
S^5/\Gamma$ where $\Gamma$ is a discrete non-abelian group.

The duality between weak and strong coupling field theories and then between
all the different superstring theories has led to a revolution in our
understanding of strings. Equally profound, is the AdS/CFT duality which is
the subject of the present article. This AdS/CFT duality is between string
theory compactified on Anti-de-Sitter space and Conformal Field Theory.

Until very recently, the possibility of testing string theory seemed at best
remote. The advent of $AdS/CFT$s and large-scale string compactification
suggest this point of view may be too pessimistic, since both could lead to $%
\sim 100TeV$ evidence for strings. With this thought in mind, we are
encouraged to build $AdS/CFT$ models with realistic fermionic structure, and
reduce to the standard model below $\sim 1TeV$.

Using AdS/CFT duality, one arrives at a class of gauge field theories of
special recent interest. The simplest compactification of a ten-dimensional
superstring on a product of an AdS space with a five-dimensional spherical
manifold leads to an ${\cal N} = 4~SU(N)$ supersymmetric gauge theory, well
known to be conformally invariant\cite{mandelstam}. By replacing the
manifold $S^5$ by an orbifold $S^5/\Gamma$ one arrives at less
supersymmetries corresponding to ${\cal N} = 2,~1 ~{\rm or}~ 0$ depending%
\cite{KS} on whether $\Gamma \subset SU(2), ~~ SU(3), ~~{\rm or} \not%
{\subset }SU(3)$ respectively, where $\Gamma$ is in all cases a subgroup of $%
SU(4) \sim SO(6)$ the isometry of the $S^5$ manifold.

It was conjectured in \cite{maldacena} that such $SU(N)$ gauge theories are
conformal in the $N \rightarrow \infty$ limit. In \cite{F1} it was
conjectured that at least a subset of the resultant nonsupersymmetric ${\cal %
N} = 0$ theories are conformal even for finite $N$. Some first steps to
check this idea were made in \cite{WS}. Model-building based on abelian $%
\Gamma$ was studied further in \cite{CV,F2,F3}, arriving in \cite{F3} at an $%
SU(3)^7$ model based on $\Gamma = Z_7$ which has three families of chiral
fermions, a correct value for ${\rm sin}^2 \theta$ and a conformal scale $%
\sim 10$~~TeV.

The case of non-abelian orbifolds bases on non-abelian $\Gamma$ has not
previously been studied, partially due to the fact that it is apparently
somewhat more mathematically sophisticated. However, we shall show here that
it can be handled equally as systematically as the abelian case and leads to
richer structures and interesting results.

In such constructions, the cancellation of chiral anomalies in the
four-dimensional theory, as is necessary in extension
of the standard model ({\it e.g.} \cite{chiral,331}),
follows from the fact that the progenitor ten-dimensional
superstring theory has cancelling hexagon anomaly\cite{hexagon}.

We consider all non-abelian discrete groups of order $g < 32$. These are
described in detail in \cite{books,FK}. There are exactly 45 such
non-abelian groups. Because the gauge group arrived at by this construction%
\cite{CV} is $\otimes_i SU(Nd_i)$ where $d_i$ are the dimensions of the
irreducible representations of $\Gamma$, one can expect to arrive at models
such as the Pati-Salam $SU(4) \times SU(2) \times SU(2)$ type\cite{PS} by
choosing $N = 2$ and combining two singlets and a doublet in the {\bf 4} of $%
SU(4)$. Indeed we shall show that such an accommodation of the standard
model is possible by using a non-abelian $\Gamma$.

The procedures for building a model within such a conformality approach are:
(1) Choose $\Gamma$; (2) Choose a proper embedding $\Gamma \subset SU(4)$ by
assigning the components of the {\bf 4} of $SU(4)$ to irreps of $\Gamma$,
while at the same time ensuring that the {\bf 6} of $SU(4)$ is real; (3)
Choose $N$, in the gauge group $\otimes_i SU(Nd_i)$. (4) Analyse the
patterns of spontaneous symmetry breaking.

In the present study we shall choose $N = 2$ and aim at the gauge group $%
SU(4) \times SU(2) \times SU(2)$. To obtain chiral fermions, it is necessary%
\cite{CV} that the {\bf 4} of $SU(4)$ be complex ${\bf 4} \neq {\bf 4}^*$.
Actually this condition is not quite sufficient to ensure chirality in the
present case because of the pseudoreality of $SU(2)$. We must ensure that
the {\bf 4} is not just pseudoreal.

This last condition means that many of our 45 candidates for $\Gamma$ do not
lead to chiral fermions. For example, $\Gamma = Q_{2n} \subset SU(2)$ has
irreps of appropriate dimensionalities for our purpose but it will not
sustain chiral fermions under $SU(4)\times SU(2) \times SU(2)$ because these
irreps are all, like $SU(2)$, pseudoreal.\footnote{%
Note that were we using $N \geq 3$ then a pseudoreal {\bf 4} would give
chiral fermions.} Applying the rule that {\bf 4} must be neither real nor
pseudoreal leaves a total of only 19 possible non-abelian discrete groups of
order $g \leq 31$. The smallest group which avoids pseudoreality has order $%
g = 16$ but gives only two families. The technical details of our systematic
search will be postponed to a future publication. Here we shall present only
the simplest interesting non-abelian case which has $g = 24$ and gives three
chiral families in a Pati-Salam-type model\cite{PS}.

Before proceeding to the details of the specific $g = 24$ case, it is worth
reminding the reader that the Conformal Field Theory (CFT) that it
exemplifies should be free of all divergences, even logarithmic ones, if the
conformality conjecture is correct, and be completely finite. Further the
theory is originating from a superstring theory in a higher-dimension (ten)
and contains gravity\cite{V,RS,GW} by compactification of the
higher-dimensional graviton already contained in that superstring theory. In
the CFT as we derive it, gravity is absent because we have not kept these
graviton modes - of course, their influence on high-energy physics
experiments is generally completely negligible unless the compactification
scale is ``large''\cite{antoniadis}; here we shall neglect the effects of
gravity.

To motivate our model it is instructive to comment on the choice of $\Gamma$
and on the choice of embedding.

If we embed only four singlets of $\Gamma$ in the {\bf 4} of $SU(4)$ then
this has the effect of abelianizing $\Gamma$ and the gauge group obtained in
the chiral sector of the theory is $SU(N)^q$. These cases can be interesting
but have already been studied\cite{CV,F2}. Thus, we require at least one
irrep of $\Gamma$ to have $d_i \geq 2$ in the embedding.

The only $\Gamma$ of order $g \leq 31$ with a {\bf 4} is $Z_5 \tilde{\times}
Z_4$ and this embedding leads to a non-chiral theory. This leaves only
embeddings with two singlets and a doublet, a triplet and a singlet or two
doublets.

The third of these choices leads to richer structures for low order $\Gamma$%
. Concentrating on them shows that of the chiral models possible, those from
groups of low order result in an insufficient number (below three) of chiral
families.

The first group that can lead to exactly three families occurs at order $g =
24$ and is $\Gamma = Z_3 \times Q$ where $Q (\equiv Q_4)$ is the group of
unit quarternions which is the smallest dicyclic group $Q_{2n}$.

There are several potential models due to the different choices for the {\bf %
4} of $SU(4)$ but only the case {\bf 4} = $(1\alpha, 1^{^{\prime}}, 2\alpha)$
leads to three families so let us describe this in some detail:

Since $Q \times Z_3$ is a direct product group, we can write the irreps as $%
R_i \otimes \alpha^{a}$ where $R_i$ is a $Q$ irrep and $\alpha^{a}$ is a $%
Z_3 $ irrep. We write $Q$ irreps as $1,~1^{^{\prime}},~1^{^{\prime\prime}},~
1^{^{\prime\prime\prime}},~2$ while the irreps of $Z_3$ are all singlets
which we call $\alpha, \alpha^2, \alpha^3 = 1$. Thus $Q \times Z_3$ has
fiveteen irreps in all and the gauge group will be of Pati-Salam type for $N
= 2$.

If we wish to break all supersymmetry, the {\bf 4} may not contain the
trivial singlet of $\Gamma$. Due to permutational symmetry among the
singlets it is sufficiently general to choose {\bf 4} = $(1%
\alpha^{a_1},~1^{^{\prime}}\alpha^{a_2},~2\alpha^{a_3})$ with $a_1 \neq 0$.

To fix the $a_i$ we note that the scalar sector of the theory which is
generated by the {\bf 6} of $SU(4)$ can be used as a constraint since the
{\bf 6} is required to be real. This leads to $a_1 + a_2 = - 2a_3 ({\rm mod}%
~3)$. Up to permutations in the chiral fermion sector the most general
choice is $a_1 = a_3= +1$ and $a_2 = 0$. Hence our choice of embedding is
\begin{equation}
{\bf 4} = (1\alpha,~1^{^{\prime}},~2\alpha)  \label{embed}
\end{equation}
with
\begin{equation}
{\bf 6} =
(1^{^{\prime}}\alpha,~2\alpha,~2\alpha^{2},~1^{^{\prime}}\alpha^{2})
\label{six}
\end{equation}
which is real as required.

We are now in a position to summarize the particle content of the theory.
The fermions are given by
\begin{equation}
\sum_I~{\bf 4}\times R_I
\end{equation}
where the $R_I$ are all the irreps of $\Gamma = Q \times Z_3$. This is:
\[
\sum_{i=1}^{3} [(2_{1}\alpha^{i},2_{2}\alpha^{i})
+(2_{3}\alpha^{i},2_{4}\alpha^{i})+(2_{2}\alpha^{i},2_{1}\alpha^{i})
+(2_{4}\alpha^{i},2_{3}\alpha^{i})+(4\alpha^{i},\overline{4}\alpha^{i})]
\]

\begin{equation}
+ \sum_{i=1}^{3} \sum_{a=1}^{4} [(2_{a}\alpha^{i},
2_{a}\alpha^{i+1})+(2_{a}\alpha^{i},4\alpha^{i+1}) + (\bar{4}%
\alpha^{i},2_{a}\alpha^{i+1})]  \label{fermions}
\end{equation}

It is convenient to represent the chiral portions of these in a quiver 
diagram.

The scalars are given by
\begin{equation}
\sum_I~{\bf 6}\times R_I
\end{equation}
and are:
\[
\sum_{i=1}^{3} \sum_{j=1(j\neq i)}^{3} [(2_{1}\alpha^{i},2_{2}
\alpha^{j})+(2_{2}\alpha^{i}, 2_{1}\alpha^{j})+(2_{3}\alpha^{i},
2_{4}\alpha^{j})+(2_{4}\alpha^{i},2_{3}\alpha^{j})
+(2_{2}\alpha^{i},2_{1}\alpha^{i})+(2_{4}\alpha^{i},2_{3}\alpha^{i})]
\]
\begin{equation}
+ \sum_{i=1}^{3} \sum_{j=1(j\neq i)}^{3} \{
\sum_{a=1}^{4}[(2_{a}\alpha^{i},4\alpha^{j}) +\bar{(4}\alpha^{i},2_{a}%
\alpha^{j} )] +(4\alpha^{i}, \bar{4}\alpha ^{i}) \}  \label{scalars}
\end{equation}
which is easily checked to be real.

The gauge group $SU(4)^3 \times SU(2)^{12}$ with chiral fermions of Eq.(\ref
{fermions}) and scalars of Eq.(\ref{scalars}) is expected to acquire
confromal invariance at an infra-red fixed point of the renormalization
group, as discussed in \cite{F1}.

To begin our examination of the symmetry breaking we first observe that if
we break the three $SU(4)$s to the totally diagonal $SU(4)$, then chirality
in the fermionic sector is lost. To avoid this we break $SU_{1}(4)$
completely and then break $SU_{\alpha }(4)\times SU_{\alpha ^{2}}(4)$ to its
diagonal subgroup $SU_{D}(4).$ The first of these steps can be achieved with
VEVs of the form $[(4_{1},2_{b}\alpha ^{k})+h.c.]$ where we are free to
choose $b$, but $k$ must be $1$ or $2$ since there are no $%
(4_{1},2_{b}\alpha ^{k=0})$ scalars. The second step requires an

$SU_{D}(4)$ singlet VEV from ($\overline{4}_{\alpha }$,4$_{\alpha^{2}})$
and/or (4$_{\alpha }$, $\overline{4}_{\alpha ^{2}})$. Once we make a choice
for $b$ (we take $b=4$), the remaining chiral fermions are, in an intuitive
notation:

\bigskip

\noindent $\ \sum_{a=1}^{3}\left[ (2_{a}\alpha \ ,1,4_{D})+(1,2_{a}\alpha
^{-1},\overline{4_{D}})\right] $

\bigskip

\noindent which has the same content as as a three family Pati-Salam model,
though with a separate $SU_{L}(2)\times SU_{R}(2)$ per family.

To further reduce the symmetry we must arrange to break to a single $%
SU_{L}(2)$ and a single $SU_{R}(2).$ This is achieved by modifying step one
where $SU_{1}(4)$ was broken. Consider the block diagonal decomposition of $%
SU_{1}(4)$ into $SU_{1L}(2) \times SU_{1R}(2).$ The representations $%
(2_{a}\alpha ,4_{1})$ and $(2_{a}\alpha ^{-1},4_{1})$ then decompose as

$(2_{a}\alpha ,4_{1})\rightarrow (2_{a}\alpha ,2,1)+(2_{a}\alpha ,1,2)$ and $%
(2_{a}\alpha ^{-1},4_{1})\rightarrow (2_{a}\alpha ^{-1},,2,1)+(2_{a}\alpha
^{-1},1,2)$. Now if we give $VEVs$ of equal magnitude to the $(2_{a}\alpha
,,2,1),$ $a=1,2,3$, and equal magnitude $VEVs$ to the $(2_{a}\alpha
^{-1},1,2)$ $a=1,2,3,$ we break $SU_{1L}(2) \times
\prod\limits_{a=1}^{3}SU(2_{a}\alpha )$ to a single $SU_{L}(2)$ and we break
$SU_{1R}(2) \times \prod\limits_{a=1}^{3}SU(2_{a}\alpha )$ to a single $%
SU_{R}(2).$ Finally, $VEVs$ for $(2_{4}\alpha ,2,1)\ $and $(2_{4}\alpha
,1,2) $ as well as $(2_{4}\alpha ^{-1},2,1)\ $and $(2_{4}\alpha ^{-1},1,2)$
insures that both $SU(2_{4}\alpha )$ and $SU(2_{4}\alpha ^{-1})$ are broken
and that only three families remain chiral. The final set of chiral fermions
is then $3[(2,1,4)+(1,2,\bar{4})]$ with gauge symmetry $SU_{L}(2) \times
SU_{R}(2) \times SU_{D}(4).$

To achieve the final reduction to the standard model, an adjoint VEV\ from ($%
\overline{4}_{\alpha }$,4$_{\alpha ^{2}})$ and/or (4$_{\alpha }$,$\overline{4%
}_{\alpha ^{2}})$ is used to break $SU_{D}(4)$ to the $SU(3)\times U(1),$
and a right handed doublet is used to break $SU_{R}(2).$

While this completes our analysis of symmetry breaking, it is worthwhile
noting the degree of constraint imposed on the symmetry and particle content
of a model as the number of irreps $N_{R}$ of the discrete group $\Gamma $
associated with the choice of orbifold changes. The number of guage groups
grows linearly in $N_{R}$, the number of scalar irreps grows roughly
quadratically with $N_{R}$, and the chiral fermion content is highly $\Gamma
$ dependent. If we require the minimal $\Gamma $ that is large enough for
the model generated to contain the fermions of the standard model and have
sufficient scalars to break the symmetry to that of the standard model, then
$\Gamma = Q \times Z_{3}$ appears to be that minimal choice\cite{FK2}.

Although a decade ago the chances of testing string theory seemed at best
remote, recent progress has given us hope that such tests may indeed be
possible in AdS/CFTs. The model provided here demonstrates the standard
model can be accomodated in these theories and suggests the possibility of a
rich spectrum of new physics just around the TeV corner.

\bigskip 
\bigskip 
\bigskip

\newpage

\bigskip

\section{Gauge unification}

\bigskip

Perhaps the most encouraging aspect of conformality is that it can replace
the traditional SusyGUT unification of $\alpha_3, \alpha_2$ and $\alpha_1$
by a group-theoretical relationship already at the TeV scale. Above this
scale there is only one independent couplig which does not
run.

\bigskip

In abelian orbifolds, the simplest three-family model occurs for $Z_7$
if we insist on ${\cal N} = 0$ for supersymmetry. As shown in \cite{F2}
this leads to the value $\sin^2 \theta = 3/13 = 0.231$ in
excellent agreement with experiment.

In this case the gauge group at the conformality scale is $(SU(3))^7$
and one of these $SU(3)$ is the color of QCD. $SU(2)_L$ is in 
a diagonal subgroup of two factors and the correctly-normalized
$U(1)_Y$ is the remaining four factors.

This choice is not arbitrary because $Z_7$ is the smallest abelian
group giving ${\cal N} = 0$ and allowing survival of three families as well as
an adeqyate scalar sector to allow spontaneous symmetry breaking
to the standard $SU(3) \times SU(2) \times U(1)$ gauge group.

Here the breaking to the standard model is most easily seen by
the intermediate step of trinification with unifying semi-simple
gauge group $SU(3)_C \times SU(3)_L \times SU(3)_R$.

\bigskip
\bigskip

More recently, with Mohapatra and Suh, we have shown \cite{FMS} how for
the non-abelian example of \cite{FK2} we
arrive at $\sin^2 \theta = 0.227$ within the context
of the left-right model. This is again supportive of our approach.

\bigskip

In the non-abelain case the conformality gauge group
is $SU(4)^3 \times SU(2)^{12}$ and is broken to the Pati-Salam
group $SU(4)_C \times SU(2)_L \times SU(2)_R$ as explained in \cite{FK2}.
It is remarkable that the $\sin^2\theta$ value comes out
so close to experiment for both $SU(3)_C \times SU(3)_L \times SU(3)_R$
and $SU(4)_C \times SU(2)_L \times SU(2)_R$.

\bigskip

Some comments on the neutrino mass spectrum are in \cite{FMS}. 

\bigskip

Of course, the fermion hierarchy for quark and lepton masses remains
challenging and it will be interesting to find whether on pushing the
present approach harder any light can be shed on this - at the moment
there is merely sufficient freedom in identification of the physical fermions
that an arbitrary mass matrix can be accommodated. To proceed further 
will require a more precise identication of the physical states.

\newpage

\bigskip

\section{Summary}

\bigskip

We have shown how $AdS/CFT$ duality leads to a large
class of models which can provide interesting extensions of the standard model of
particle phenomenology. The naturally occurring ${\cal N} = 4$ extended
supersymmetry was completely broken to
${\cal N} = 0$ by choice of orbifolds $S^5/\Gamma$ such that
$\Gamma \not\subset SU(3)$.

\bigskip

In \cite{FKlong}, we studied systematically all such non-abelian $\Gamma$
with order $g \leq 31$. We show how chiral fermions require that the
embedding of $\Gamma$ be neither real nor pseudoreal. This reduces dramatically
the number of possibilities to obtain chiral fermions.
Nevertheless, many candidates for models which contain the
chiral fermions of the three-family standard model were found.

However, the requirement that the spontaneous symmetry breaking down to the correct gauge symmetry
of the standard model be permitted by the prescribed scalar representations
eliminates most of the surviving models. We found only one allowed model
based on the $\Gamma = 24/7$ orbifold. We had initially expected to find more examples in our search.
The moral for model-building is interesting. Without the rigid framework of string duality
the scalar sector would be arbitrarily chosen to
permit the required spontaneous symmetry breaking. This is the normal practice in the standard model,
in grand unification, in supersymmetry and so on.
With string duality, the scalar sector is prescribed by the construction
and only in one very special case does it permit the required symmetry breaking.

\bigskip

This leads us to give more credence to the $\Gamma = 24/7$ example that does work
and to encourage its further study to check whether it can have any connection
to the real world.

\bigskip
\bigskip
\bigskip

\section*{Acknowledgement}

It is a pleasure to thank W.F. Shively, C. Vafa, T.W. Kephart, R.N. Mohapatra
and S. Suh for their collaboration at stages of this
ongoing attempt to connect M theory and the real world. 
This work was supported in part by the US Department of Energy
under Grant No. DE-FG02-97ER-41036.

\end{document}